\documentclass{ws-p8-50x6-00}
\usepackage{graphics}

\def\be{\begin{equation}}
\def\ee{\end{equation}}
\def\bea{\begin{eqnarray}}
\def\eea{\end{eqnarray}}

\begin{document}

\title{Study of gluon fragmentation and colour octet neutralization in DELPHI}

\author{B.Buschbeck and \underline{F.Mandl}}

\address{Institute for High Energy Physics of the OAW, Nikolsdorferstrasze 18,
\\A-1050 Wien, Austria\\E-mail:mandl@hephy.oeaw.ac.at}

\maketitle
\abstracts{Using the full statistics of the DELPHI experiment
at $\sqrt{s}=91~GeV$ 3-jet events are selected and
gluon respectively quark jet enriched subsamples are defined.
The leading systems of the two kinds of jets are determined
using rapidity gaps. The sum of charges of the leading systems is studied.
It is found that for gluon-jets there is a significant excess of
leading systems with total charge zero when compared to Monte Carlo simulations
with JETSET. The corresponding leading systems of quark-jets do not
exhibit such an excess.
The mass spectra of the
leading systems with total charge zero are studied.
}

\section{Introduction}
The study of the leading hadrons in gluon jets may give interesting insight
into the mechanism of colour neutralization.
In principle there could exist a direct neutralization
of the colour octet field of the fast gluon by the creation of gluon pairs.
This mechanism
is called `colour octet neutralization' \cite{ochs1}. As a
consequence bound states with valence gluons
(glueballs and gluonic mesons) might be
eventually produced and observed.
The existence of gluonic states is
predicted by non-perturbative QCD and has been searched for since more than
20 years in different experiments \cite{pred}.
The very existence of such states, however, is not yet
established.

The present study is following
a suggestion of Minkowski and Ochs \cite{ochs1,ochs2}
to search for colour octet neutralization (and for
glueballs and gluonic mesons) in gluon jets, produced in 3-jet events
in $e^+e^-$ reactions,
by defining a leading system
followed by a rapidity gap $\Delta y$
empty of hadrons.
The leading systems of both gluon- and quark-jets 
are then compared 
to the predictions of the Monte Carlo model JETSET which is
not including the mechanism of octet neutralization.
The price to pay for a selection of jets with a rapidity gap 
is however a strong reduction
of the number of events because of the Sudakov form
factor \cite{suda}.
The sum of charges (SQ) in the leading particle system of gluon jets is 
compared with the resp. Monte
Carlo prediction to search for a surplus of events with SQ=0.
Furthermore the same investigation for quark jets should not result in a
surplus of any charge.

In ref. \cite{etal3} it has been observed that there were
no significant deviations from Monte Carlo predictions
for resonance production - in particular that of the $\eta$
in quark- and gluon-jets.
Therefore one can expect that octet neutralization is a relatively rare
process - if it exists at all.
\vspace{-0.2cm}
\section{Data sample and 3-jet event selection}
The data sample used has been taken by the DELPHI experiment at the
LEP collider at $\sqrt{s}=91~GeV$ in the years 1992-1995.
About 230000 3-jet-event have been selected,
which have been obtained
by using the appropriate cuts for track quality
and for the hadronic event type \cite{delcut} as well as applying a $k_t$
cluster algorithm
(Durham) \cite{durham} with $y_{cut}=0.015$.
For the jet determination all topologies have been used with
$\Theta_2 , \Theta_3  =  135^o \pm 35^o$, where the jets had been numbered
with respect to the calculated energy i.e. $E_3 \leq E_2 \leq E_1$ and
with e.g. $\Theta_2$ being
the angle between the first and third jet (so called "asymmetric events")
\cite{x1,x2,x3}.
The jet with the highest energy $E_1$ (jet-1) is in most cases a quark-jet,
that with
the smallest energy $E_3$ (jet-3) the gluon jet. Monte Carlo simulations show for
the above mentioned conditions for jet-1 a quark-jet contribution of $\geq 90$\%
and for jet-3 a gluon-jet contribution of about 70\%.
This is e.g. in agreement with the
numbers quoted by L3 with similar jet energies and selections
\cite{l3gluon}
Heavy quark
(b- and c- quarks) events are classified using an impact parameter technique
\cite{boris,klapp}. In the present study events are only accepted
if they do not exhibit a b-quark signal. The intention is to compare gluon
jets only to 'light-quark' jets.
  A corresponding sample of Monte Carlo simulations
(JETSET) of about
twice the event- statistics has been created for comparisons.
All comparisons are done at the detector level.

\section{Preliminary Results for 3-Jet Events}
\subsection{Sum of Charges in the Leading System with Rapidity Gap}
After the selection of 3-jet events and the determination
of enriched quark and gluon jet samples
the leading hadronic system of the jet is singled out
by requiring a rapidity interval ($\Delta y \ge 2$) between the particles
(charged and
neutral) belonging to this system and the rest
of hadrons produced in either kind of jet.
For the charged particles the momenta are required to be larger than
0.2~GeV, for the neutrals this requirement is 0.5~GeV.
The requirement of the rapidity interval $\Delta y \ge 2$ below the
leading system to be empty of hadrons
reduces the number of events to only 3$\%$.
\begin{figure}
\vspace{5.5cm}
\begin{picture}(20,140)(0,0)
\put(-20,0){\mbox{\resizebox{55mm}{!}{\includegraphics{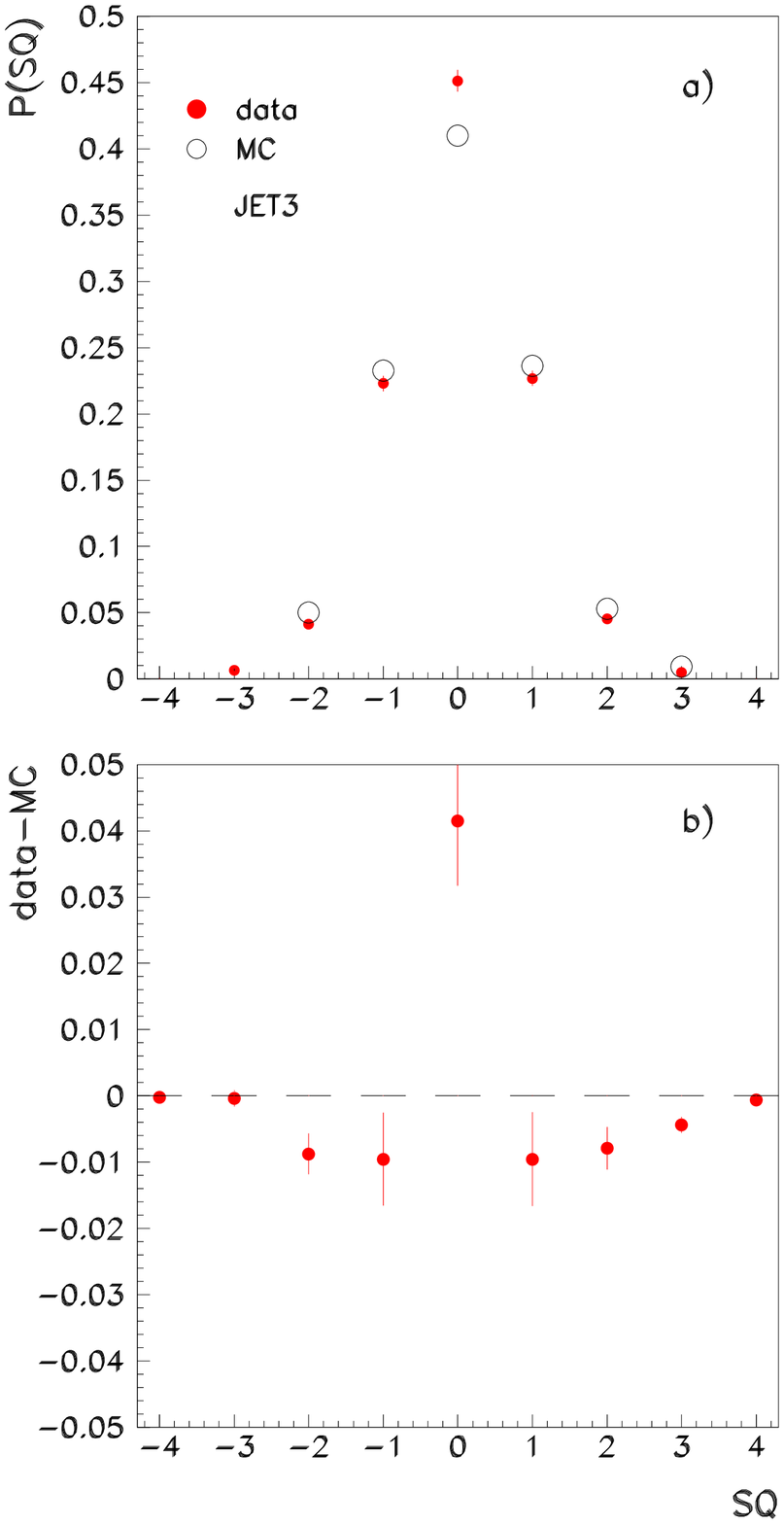}}}}
\put(0,-10){\makebox(0,0)[l]{Fig.1  Sum of charges, leading}}
\put(23,-21){\makebox(0,0)[l]{system, gluon-jets}}
\end{picture}
\end{figure}

\begin{figure}
\vspace{-0.6cm} \hspace{5.8cm}
\begin{picture}(0,50)(0,0)
\put(0,50){\mbox{\resizebox{55mm}{!}{\includegraphics{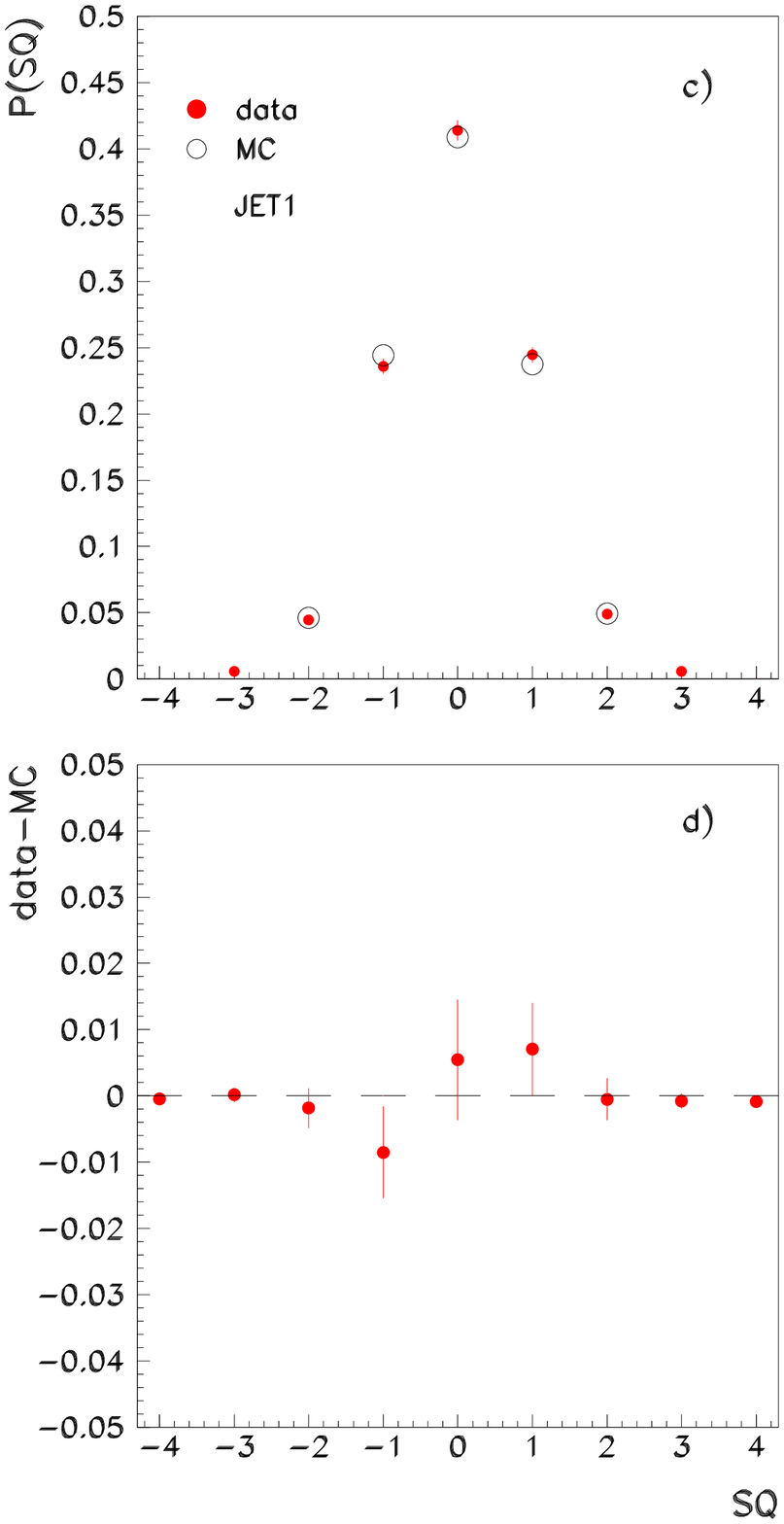}}}}
\put(20,40){\makebox(0,0)[l]{Fig.2   Sum of charges, leading}}
\put(43,29){\makebox(0,0)[l]{system, quark-jets}}
\end{picture}
\vspace{-1.6cm}
\end{figure}
The sum of charges of the
particles belonging to the leading system defined as above is given
in Fig.1 for gluon-jets and in Fig.2 for quark-jets (full circles) and
compared to JETSET Monte Carlo simulations (open circles).
The numbers P(SQ) in the upper plots are defined
as the number of events (or Monte Carlo events) with a certain 
SQ divided by the total number of selected
$\Delta y \ge 2$ events - (or the corresponding
Monte Carlo events). They are therefore an estimate for the
probability of an event to have a certain SQ.
\begin{figure}
\mbox{\epsfig{file=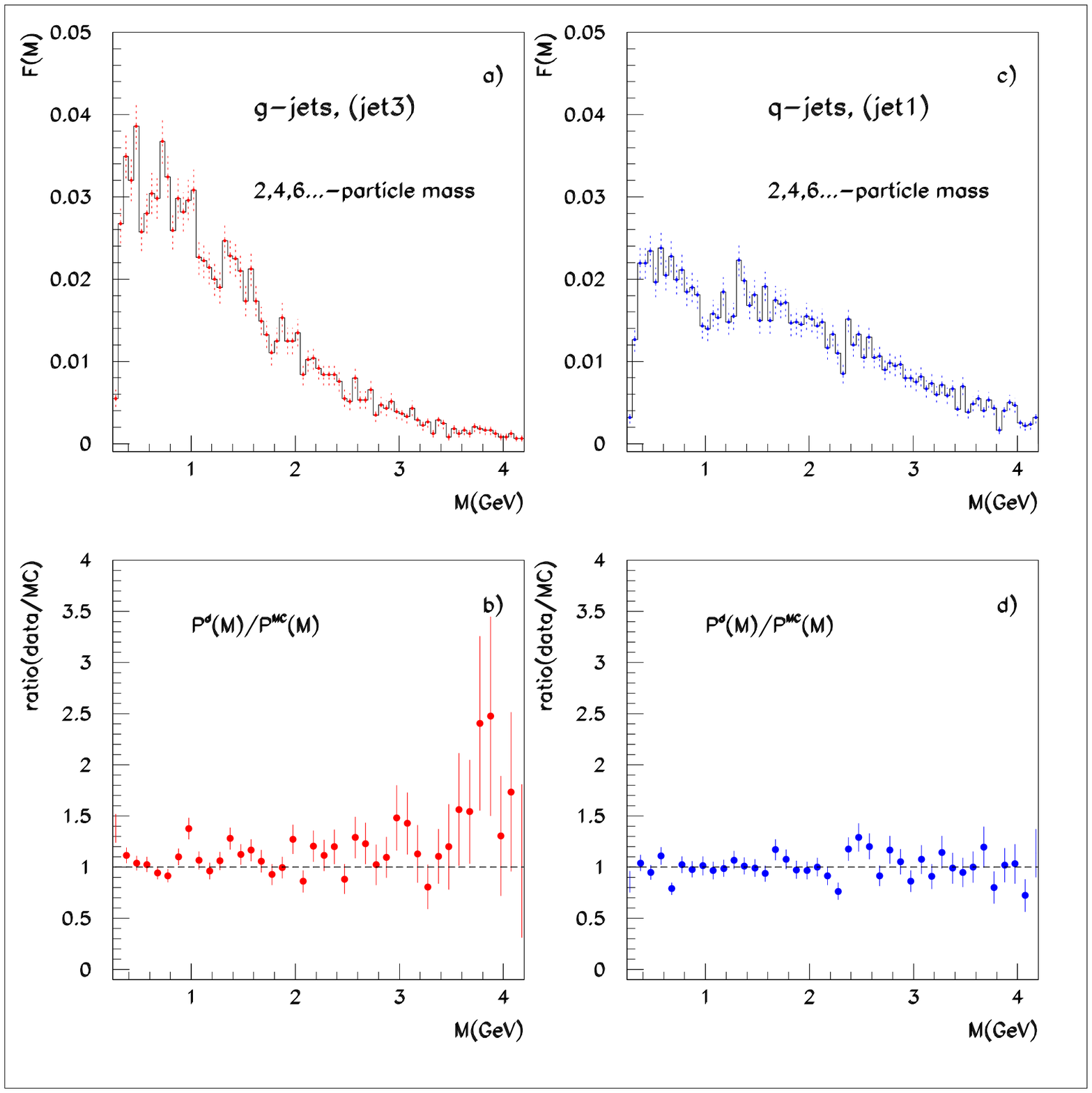,width=12.1cm}}
\put(-325,-10){\makebox(0,0)[l]{Fig.3  Effective mass distribution of the leading system for both }}
\put(-302,-21){\makebox(0,0)[l]{gluon-jets (a) and quark-jets (c) as well as the respective ratios}}
\put(-302,-32){\makebox(0,0)[l]{ata/JETSET Monte Carlo simulations (b resp. c)}}
\vspace{-0.1cm}
\end{figure}
 The SQ distributions  of the leading system for the {\em gluon-jet}
(Fig.1) shows,
for SQ=0, a
{\em striking difference}
between data and simulation. There is a significant enhancement of the
SQ of the data at SQ=0 as expected (see Section 1, ref. \cite{ochs1,ochs2})
when the process of colour octet neutralization is present
(it is, as mentioned above,
absent in the simulation).
On the other hand, there is {\em no significant difference} of the
SQ of the
leading system, for SQ=0, between data and JETSET
simulation in the case of {\em quark-jets} (Fig.2).
The lower parts of Figs. 2 and 3 show the difference of the P(SQ) between
the data and the JETSET Monte Carlo simulation.
This difference amounts, for the gluon-jet (Fig.1) to about $9\%$,
that is more than 4 standard deviations from zero, for the quark-jet (Fig.2)
this difference is compatible with zero !
\subsection{Mass Spectra}
\noindent Fig.3 shows, for both the gluon-jet and the quark-jet, the effective
mass distribution $M$
of the leading system, no
neutrals included, with
rapidity gap required for the charged particles $\Delta y \ge 2$ and the
total charge of the system being zero. The number of charged particles in the
leading system has to be 2,4,6 etc.
Several peaks can be observed for the {\em gluon-jet} (Fig.3a). 
One peak around $M \sim 0.8~GeV$
might be attributed to the $\rho$ resonance, another at
$M \le 0.5~GeV$ to a reflection of $\eta$, $\eta^`$ and $\omega$.
Other peaks ($M \le 1~GeV$, $M \sim 1.4~GeV$, ...) are not
understood yet and their statistical significance is weak.
The ratio of the $M$ distributions of the leading system
data/simulation is given in Fig.3b.
The $\rho$ peak
and the enhancement at
$M \le 0.5~GeV$ seem to be understood by the simulation.
Besides the region at very low $M \le 0.3~GeV$, the 2 enhancements
in the $M$ distributions of the leading system of the gluon-jet
at $M \le 1~GeV$ and $M \sim 1.4~GeV$
are not reproduced by the JETSET Monte Carlo simulation.

Fig.3c shows the corresponding mass distribution of {\em quark-jets} and
Fig.3d shows the ratio of the $M$ distributions of the leading system
data/simulation for the quark-jet.
Compared to the gluon-jet there is no
enhancement at $M \le 1~GeV$ in Fig.3c and in Fig.3d
no excess compared to the simulation 
at $M \sim 1.4~GeV$.

Since the effect of the neutrals are not well understood yet ($\gamma$'s,
$\pi^0$'s, and neutral decay modes of $\eta$'s, $\eta^`$'s,
$\omega$'s etc.) and because of the limited
statistics no conclusions can be drawn yet from the mass distributions.

\section{Summary and Conclusions}
In the present study first efforts have been undertaken to search for the
existence of octet neutralization in the fragmentation of gluon-jets.
The full statistics of 1992-1995 at $\sqrt{s}=91.1~GeV$ obtained
by the DELPHI collaboration is used to select 3-jet events and to single out
thereof quark-jets (purity $\ge 90\%$) and gluon-jets
(purity $\ge 70\%$). A leading system of a jet is defined which is separated
from the rest of the low energy particles by a rapidity gap of width
$\Delta y \ge 2$ being empty of hadrons. The sum of charges of this leading
system is studied. For the gluon-jets an enhancement of neutral leading
systems over the Monte Carlo prediction of about $9\%$ is
seen (more than 4 standart deviations above zero), on the other
hand, no such enhancement is seen
in the quark-jet !
An even more significant deviation is revealed (Fig. not shown) for the sum
of charges of the 2 fastest tracks without demanding a rapidity cut
($3.5\%$ effect but with very high significance).

In order to assess the existence of colour octet neutralization further checks
have to be done - determination of the quantum numbers of the leading
system, better separation of the gluon-jet, better insight into the
role of the neutrals etc. It can, however, be argued that there 
{\em is} an
intrinsic shortcoming for the JETSET Monte Carlo simulation describing
the sum of charges in the leading system of the  gluon-jet !

The effective mass of the leading system with sum of charges SQ = 0
is studied. An enhancement for
the gluon-jet at
$M \le 1~GeV$, 
which is not seen in
the simulation nor for the quarkjet
and another at $M \sim 1.4~GeV$
is yet of only weak
statistical significance.
\section*{Acknowledgements}
We thank W.Ochs for encouraging us to start the above study and for 
discussions in the course of it, O.Klapp for technical support for the
jet selection and M.Siebel for valuable comments.

\end{document}